\documentclass[11pt,twoside]{article}

\usepackage{asp2006}
\usepackage{epsf}
\usepackage{lscape}

\begin{document}
\title{The Large Scale Structure of LINERs and Seyferts and
Implications for their Central Engines.}  

\author{Anca Constantin \& Michael S. Vogeley}

\affil{Department of Physics, Drexel University, Philadelphia, PA
19104} 

\begin{abstract} 
We discuss here the spatial clustering of Seyferts and LINERs and
consequences for their central engines.  We show that Seyferts are
less clustered than LINERs, and that this difference is not driven by
the morphology-density relation, but it is related to the difference
in clustering as a function of level of activity in these systems and
the amount of fuel available for accretion.  LINERs, which are the
most clustered among AGN, show the lowest luminosities and obscuration
levels, and relatively low gas densities, suggesting that these
objects harbor black holes that are relatively massive yet weakly
active or inefficient in their accretion, probably due to the
insufficiency of their fuel supply.  Seyferts, which are weakly
clustered, are very luminous, show generally high gas densities and
large quantities of obscuring material, suggesting that in these
systems the black holes are less massive but abundantly fueled and
therefore accrete quickly and probably efficiently enough to clearly
dominate the ionization.


\end{abstract}



\section*{Spatial Clustering and Accretion Activity}   

The nature of the central engine of Active Galactic Nuclei (AGN) can
be independently constrained via quantifying these systems' large
scale structure.  The pattern of their spatial clustering provides
important diagnostics for the masses and number density of the dark
matter haloes in which they reside, and consequently, for the masses
of their central accreting black holes.  Therefore, comparisons of the
clustering properties of AGN of different characteristics may indicate
intrinsic differences in their ionization mechanisms and even gauge
the accretion strength and time cycle.  Such analysis can thus yield a
phenomenological link between black hole growth and galaxy formation
and evolution processes.

Many studies of AGN clustering characteristics compare passive
galaxies with AGN as a general class and do not examine the properties
of sub-populations such as Seyferts and LINERs separately.  In such
investigations, the AGN population is found to occupy a uniform
fraction of galaxies, to be common to a large range of galactic
environments, and to follow the distribution of the whole galaxy
population, and therefore to be unbiased tracers of mass in the
universe \citep{mil03, wak04}.  However, when divided by their
[\ion{O}{iii}] luminosities, AGN show different clustering properties
\citep{wak04}, and different environmental preference \citep{kau04}.
Given that Seyferts and LINERs represent quite distinct distributions
in $L_{\rm [O III]}$ \citep{kau03}, they might also cluster
differently.  To investigate these issues we have analyzed the
clustering properties of a variety of nearby actively line-emitting
galaxies by including important spectral diagnostics that probe the
astrophysics of AGN \citep{con06}.  We summarize here the results
pertaining to LINERs and Seyfert galaxies.

\section*{The Data}   

To estimate the clustering properties of Seyferts and LINERs, we use
the Sloan Digital Sky Survey (SDSS) DR2 main galaxy sample, which
includes every extended object with the apparent $r$-band magnitude
$14.5 \leq m_r \geq 17.7$, and which lacks broad line emission (i.e.,
is not a quasars).  To avoid biases caused by the strong dependence on
luminosity of the sample properties, and by the fact that only the
intrinsically bright galaxies are seen at large distances, which
creates a gradient of the number density of galaxies with distance, we
work here with absolute magnitude limited samples, with $-21.6 < M_r <
-20.2$, which correspond to a redshift range of $0.05 < z < 0.12$, or
a comoving distance range of $ 148.5 < r < 350.1~ h^{-1}$ Mpc.


We construct samples of Seyferts and LINERs by employing three
emission-line diagnostic diagrams based on constraints on line flux
ratios of: [\ion{O}{iii}]$\lambda$5007/H$\beta$,
[\ion{N}{ii}]$\lambda$6583/H$\alpha$,
[\ion{S}{ii}]$\lambda\lambda$6716,6731/H$\alpha$, and
[\ion{O}{i}]$\lambda$6300/H$\alpha$, on sources that show at least
2$-\sigma$ detection in all of these six emission-lines.  For more
details on the classification scheme see \citet{con06}.

Note that only about 5\% of all the strong line emitters are Seyferts,
and that the majority of the sources that might or clearly exhibit
accretion activity are classified as LINERs based on their line flux
ratios.  This means that samples of AGN that consider these two
classes together are severely skewed toward the LINER-like behavior,
and are not capable of offering any information specific to Seyferts.
The SDSS-based AGN definitions that leniently use only the
[\ion{O}{iii}]/H$\beta$ versus [\ion{N}{ii}]/H$\alpha$ diagnostic
diagram are particularly susceptible to these effects.


\section*{The Clustering Amplitude and the Accretion Activity}   

We measure the two-point correlation function in redshift space
$\xi(s)$ for Seyferts and LINERs.  Power-law fits $\xi(s) =
\big(s/s_0\big)^{-\gamma}$ over separations $1 < s < 10~ h^{-1}$ Mpc
yield the clustering amplitude $s_0$.  This parameter is a measure of
the strength of the clustering as it can be viewed as the scale at
which the probability of finding a galaxy pair is, in average, double
that measured within a random sample.  Our measurements of $s_0$
reveal a lower value for Seyferts ($5.7 \pm 0.6~ h^{-1}$Mpc) than for
LINERs ($7.8 \pm 0.6~ h^{-1}$Mpc), whose spatial distribution seems to
follow that of the average galaxies ($s_0 = 7.8 \pm 0.5~ h^{-1}$Mpc
for the whole sample of galaxies).  This suggests that LINERs inhabit
denser environments than Seyferts.

Moreover, this difference in $s_0$ points to a more fundamental
contrast between Seyferts and LINERs, i.e., in the mass of their
central accreting black holes.  If structure formation is
hierarchical, then the rarest and most massive virialized dark matter
halos cluster the most strongly \citep{kai84}.  Therefore, the
different $s_0$ values indicate that Seyferts reside in less massive
haloes than those harboring galaxies with LINER-like activity in their
centers.  Consequently, given that the mass of the dark matter haloes
may correlate with that of the black holes that lie at the centers of
their resident galaxies \citep{fer02, bae03}, this finding suggests
that Seyferts' black holes are less massive than those of LINERs.


An important thing to investigate is whether this discrepancy in the
clustering properties of Seyferts and LINERs derives from their host
morphologies, as we know that the red, early type of galaxies are more
clustered than the blue, late types ones, i.e., the morphology-density
relation.  We have thus compared their host properties in terms of
both the concentration index $C$ (a relatively good proxy for their
morphological type) and the $u-r$ color, and found that the LINERs and
Seyfert distributions in both of these two parameters are consistent
with having the same parent populations (Kolmogorov-Smirnov
probabilities are: KS$_{S-L}^{~~C}$ = 0.104, KS$_{S-L}^{~~u-r}$ =
0.617).  This analysis clearly shows that the different clustering
that Seyferts and LINERs exhibit is not driven by the
morphology-density relation.


The difference in the clustering amplitude of Seyferts and LINERs is
however closely related to the difference in clustering as a function
of accretion activity in these systems.  We found that parameters like
the [\ion{O}{i}] and [\ion{O}{iii}] luminosities, that can be used as
proxies for the fueling rate, have a very strong effect on the
clustering amplitude.  Objects with low luminosities in both of these
lines are very strongly clustered, while their dimmer counterparts
remain weakly clustered.  The difference in $s_0$ in both such cases
is about $5-\sigma$.  Interestingly, when we separate out the low and
high $L_{\rm [O I]}$ and $L_{\rm [O III]}$ sources into spectral
classes, we find that the low luminosity objects are basically the
LINERs, and that, as expected, the weakly clustered high $L_{\rm [O
I]}$ and $L_{\rm [O III]}$ systems include the majority of Seyferts.

Moreover, the amount of fuel, as gauged by the amount of obscuration
(as measured by the neutral hydrogen column density $N_H$ derived from
the H$\alpha$/H$\beta$ Balmer decrements) and the gas densities
($n_e$, estimated from the [\ion{S}{ii}] ratios), seems to also
correlate with the strength of accretion and $s_0$.  LINERs and
Seyferts are again at the extreme ends in the distributions in their
$N_H$ and $n_e$, with LINERs exhibiting particularly low levels of
obscuration and $n_e$ while Seyfert galaxies show generally high
densities and moderately high $N_H$.

\section*{Results and Possible Interpretation}

A brief overview of our main results and a possible interpretation is
given in Table ~\ref{tbl-summary}, and we list our conclusions as
follows.

\begin{table}[!ht]
\caption{Summary of Results and Interpretation}
\smallskip
\begin{center}
\label{tbl-summary}
{\small
\begin{tabular}{lll}
\tableline
\noalign{\smallskip}
 & \multicolumn{2}{c}{Sample}\\
\noalign{\smallskip}
\cline{2-3}
\noalign{\smallskip}
Property  &    Seyferts &  LINERs \\
\noalign{\smallskip}
\tableline
\noalign{\smallskip}

clustering ($s_0$)...................... & weak & strong ({\it like galaxies})\\
fueling rate ($L_{\rm [O I]}$, $L_{\rm [O III]}$)... &relatively high ({\it efficient?})& (very) low ({\it inefficient?})\\
fuel available ($n_e$)................&  high        & moderately high \\
 ~~~~                    &              & ({\it wide range})\\
obscuration..........................  &  wide range  & low\\
host morphology..................&redish, earlier type & redish, earlier type \\
                  &                     & ({\it like Seyferts})\\
$M_{\rm BH}$....................................  &  small       & large\\
life-time...............................   & short        & long \\

\noalign{\smallskip}
\tableline
\end{tabular}
}
\end{center}
\end{table}

(i) We found that, contrary to conclusions of previous studies, AGN in
the local universe are not clustered like galaxies.  Only LINERs
cluster like normal galaxies.  Seyferts are significantly less
clustered than the general galaxy population.

(ii) Host galaxy properties and, consequently, the morphology-density
relation, do not strongly influence the difference in the spatial
clustering found between Seyferts and LINERs. The difference in $s_0$
is however strongly related to the level of activity in these nuclei,
as suggested by their [\ion{O}{iii}] and [\ion{O}{i}] line
luminosities.

(iii) The picture that emerges from these results is that Seyferts are
the result of relatively active but small black holes, while LINERs
are the manifestation of weakly accreting but more massive ones.  It
is then probably the case that Seyferts live in their active state for
only short periods of time, which is consistent with the fact that we
only see a small numbers of such systems, while LINERs can persist in
their low state for much longer, and are therefore more ubiquitous.

(iv) Hints for what may cause differences in the nuclear activity in
the low luminosity AGN, or simply its detection, come from the fact
that both $n_e$ and $N_H$ are significantly different in LINERs and
Seyferts.  A possible cause for the feeble activity in LINERs may be
that fuel is not as abundant as in the more active systems.

\vspace{0.25cm} Previous (and yet many of the present) studies of AGN
clustering and other statistical investigations of their properties
have not considered separating out different types of AGN.  However,
as our analysis clearly shows it, Seyferts and LINERs remain distinct
in their behavior, and should be considered separately.  A potential
idea to explore in the future is that these systems might be the
manifestation of different stages in an evolutionary sequence in the
AGN lifetime.  Our recent study of AGN activity in the most underdense
regions of the universe, the voids, provides some evidence for this
scenario \citep{con07}.

\acknowledgements 
AC thanks the organizers for invitation and hospitality. Support for
this work was provided by NASA through grant NAG5-12243 and NSF grant
AST-0507647.


\end{document}